\documentclass[twocolumn,
aps,nofootinbib,showpacs,showkeys,preprint
tightenlines,preprintnumbers,superscriptaddress]{revtex4}

\usepackage{epsf,epsfig,subfigure,graphicx,amsmath,amssymb}
\usepackage{color}

\newcommand{\gev}{\,\textrm{GeV}}

\newcommand{\Mp}{M_{\rm P}}

\newcommand{\Mg}{$M_{\rm GUT}$}

\newcommand{\ie}{{\it i.e.~}}
\newcommand{\etal}{{\it et al.}\,}

\begin{document}

\title{Perspective on completing natural inflation}
\author{Ki-Young Choi}
\address{Korea Astronomy and Space Science Institute, Daejeon 305-348, Republic of  Korea, }
\author{Jihn E. Kim}
\address{ Department of Physics, Seoul National University, Seoul 151-747, Republic of Korea, and \\
    Department of Physics, Kyung Hee University, Seoul 130-701,  Republic of Korea,
 }
   \author{Bumseok Kyae}
\address{ Department of Physics, Pusan National University, Busan 609-735,  Republic of Korea
 }

\begin{abstract}
We present a perspective on the inflation paths in $2-, 3-, \cdots, N-$flation models. The number of non-Abelian gauge groups for a natural inflation is restricted in string compactification, and we argue that the most plausible completion of natural inflation from a theory perspective  is the 2--flation.
\end{abstract}


\maketitle

%%%%%%%%%%%%%%%%%%%%%%%%%%%%%%%%%%%%%%%%%%%%%%%%%%%%%%%%%%%%%%%%%%%%%
\section{Introduction}\label{sec:Introduction}

Completing natural inflation has attracted a great deal of attention \cite{natural14} after the BICEP2 result \cite{BICEP14}. The idea was presented some time ago \cite{KNP05}.

Cosmic inflation is an attractive paradigm for a solution of the homogeneity
and flatness problems \cite{inflationold, Infnew1,InfNew2}.   For a sufficient inflation with the e-fold number $e>70$,\footnote{The number of e-foldings required in inflation depends on the specific models  as well as the dynamics after inflation. Even though the number relevant for the observed CMB anisotropies is typically around 50--60, here we use the minimum value 70 given in Ref. \cite{Tsusi14} sufficient for most of inflationary models. }
 one needs small slow-roll inflation-parameters, $\epsilon\,(\equiv\frac12\Mp^2( V'/V)^2)$ and $\eta\,(\equiv\Mp^2 V''/V)$ \cite{MukhanovBk,lyth99}. Single bubble inflation was proposed with the initial condition near the origin in the Coleman-Weinberg type logarithmically-flat hilltop potential \cite{ColemanW73}, or at a large field value for a chaotic type potential \cite{Linde83}.
With the slow-roll conditions satisfied, the local non-Gaussianities
$|f_{\rm NL}^{\rm local}|$ are much smaller than 1 for a single field inflation \cite{Kuro}, which was observed by the Planck 2013 data \cite{Planckinf}.
In addition, the hybrid inflation predicting $n_s>1$ (arising from the hilltop inflation) \cite{KimHilltop14} and the  $\lambda \phi^4$ chaotic inflation are disfavored from the data \cite{Planckinf}.

The negligible non-Gaussianity pin down the inflation models to  the single field  $m^2 \phi^2$ chaotic inflation \cite{natural14} or the multi-field hilltop inflation \cite{KimHilltop14}.  The  $m^2 \phi^2$ chaotic inflation needs a fine-tuning of order $m^2\approx 10^{-10}$ in units of the reduced Planck mass, $\Mp\simeq 2.44\times 10^{18\,}\gev$.  For the predictability of the Einstein equation, we need that the potential $V$ during inflation must be much smaller than $\Mp^4$. In fact, this can be easily realized in {\it natural inflation} where there exists a GUT scale heavy axion coupling to a GUT scale confining force \cite{Freese90}.
With the heavy axion potential at the GUT scale ($\approx\Lambda_{\rm GUT}\approx$\Mg),
the explicit breaking potential of the Peccei-Quinn (PQ) symmetry  is given by
$\propto\frac12\Lambda^4_{\rm GUT}(1-\cos(a/f))$; thus the potential energy is bounded by $\Lambda^4_{\rm GUT}$.

The  $m^2 \phi^2$ chaotic inflation has a problem, ``Why does one keep only the quadratic term?''~ It is known that a large trans-Planckian field value is needed in the  $m^2 \phi^2$ chaotic inflation for a large tensor-to-scalar ratio $r$, which is known as the Lyth bound $\langle\phi \rangle>15\,\Mp$ \cite{Lyth97}.  In particular, with the large trans-Planckian field value higher order terms might be more important  \cite{KimHilltop14}.
To reconcile the trans-Planckian field value with the natural inflation idea, Kim, Nilles, and Peloso (KNP) introduced two axions and two confining forces at the GUT scale. It has been generalized to N-flation \cite{Nflation}.

An ultra-violet completed theory, in particular the heterotic string theory, may not allow a large number of non-Abelian gauge groups. We scrutinize the inflaton path, arising from the limited rank of the total gauge group, and present an argument that 2-flation, \ie the
KNP type, is an easily realizable one.

In Sec. \ref{sec:KNP}, we briefly review the KNP scenario and its N-flation extension.
In Sec. \ref{sec:numberG}, we discuss the maximum rank of the heterotic string, which is argued for a limitation of the number of GUT scale confining gauge groups.
Sec. \ref{sec:Conclusion} is a conclusion.

%%%%%%%%%%%%%%%%%%%%%%%%%%%%%%%%%%%%%%%%%%%%%%%%%%%%%%%%%%%%%%%%%%%%%%%%%%%%%%%
\section{The 2-flation} \label{sec:KNP}

A large vacuum expectation value (VEV) of a scalar field is possible with a small mass parameter if a very small coupling constant $\lambda$ is assumed,
\begin{equation}
V=\frac14\lambda(|\phi|^2-f^2)^2.\label{eq:HilltopV}
\end{equation}
The mass parameter in this theory is $m^2=\lambda f^2$. With a GUT scale $m$, $f$ can be trans-Planckian of order $>10\Mp$ for $\lambda< 10^{-6}$. However, the potential (\ref{eq:HilltopV}) with the small $\lambda$ describes inflation starting from near the {\it convex} hilltop point (due to the high temperature effect before inflation) and hence it is not favored by the BICEP2 data \cite{KimHilltop14}.
This has led to the recent surge of studies on {\it concave} potentials near the origin of the field space in case of single field inflations \cite{natural14}. The concave potentials give positive $\eta$'s.

The simplest concave potential is the $m^2\phi^2$ chaotic potential. Since this potential is not bounded from above, the natural inflation with a GUT scale confining force has been introduced so that the potential is bounded from above \cite{Freese90} where the GUT scale axion is the inflaton and the inflaton potential is
\begin{equation}
V=\Lambda_{GUT}^4\left(1-\cos\frac{a_N}{f_N}\right).\label{eq:Freese}
\end{equation}
With O(1) parameters at the GUT scale, $f_N$ is O(\Mg).
 One may argue that the potential  (\ref{eq:Freese}) is in the angle direction, and
a large $f_N$ can result, using the radial-direction potential (\ref{eq:HilltopV}), with a very small  $\lambda\,(< 10^{-6})$. However, the potential (\ref{eq:HilltopV}) with the small $\lambda$ already describes inflation starting from near the {\it convex} hilltop point (due to the high temperature effect before inflation), and hence it is not favored by the BICEP2 data \cite{KimHilltop14}. For the radial direction to roll down quickly, we need $\lambda\gg 10^{-6}$ and $f_N$ is of order \Mg.

Since we need a trans-Planckian value for the decay constant of the GUT axion,
the KNP model has been proposed with two axions $a_1$ and $a_2$ and two GUT scale ($\Lambda_1$ and $\Lambda_2$) confining forces, resulting in the following minus-cosine potentials
\begin{eqnarray}
V&=&\Lambda_{1}^4\left(1-\cos\left[\alpha\frac{a_1}{f_1}+\beta\frac{a_2}{f_2}\right]\right)
\nonumber\\
\quad&+& \Lambda_{2}^4\left(1-\cos\left[\gamma\frac{a_1}{f_1}+\delta\frac{a_2}{f_2}\right]\right),\label{eq:KNP2}
\end{eqnarray}
where $\alpha,\beta,\gamma$, and $\delta$ are determined by two U(1) quantum numbers.
 Of course, $f_1$ and $f_2$ are O(\Mg).
Let us comment a few issues related to the above potentials.

%%%%%%%%%%%%%%%%%%%%%%%%%%%%%%%%%%%%%%%%
\subsection{One confining force}\label{subsec:one}
\begin{figure}
\begin{center}
\includegraphics[width=3.5cm]{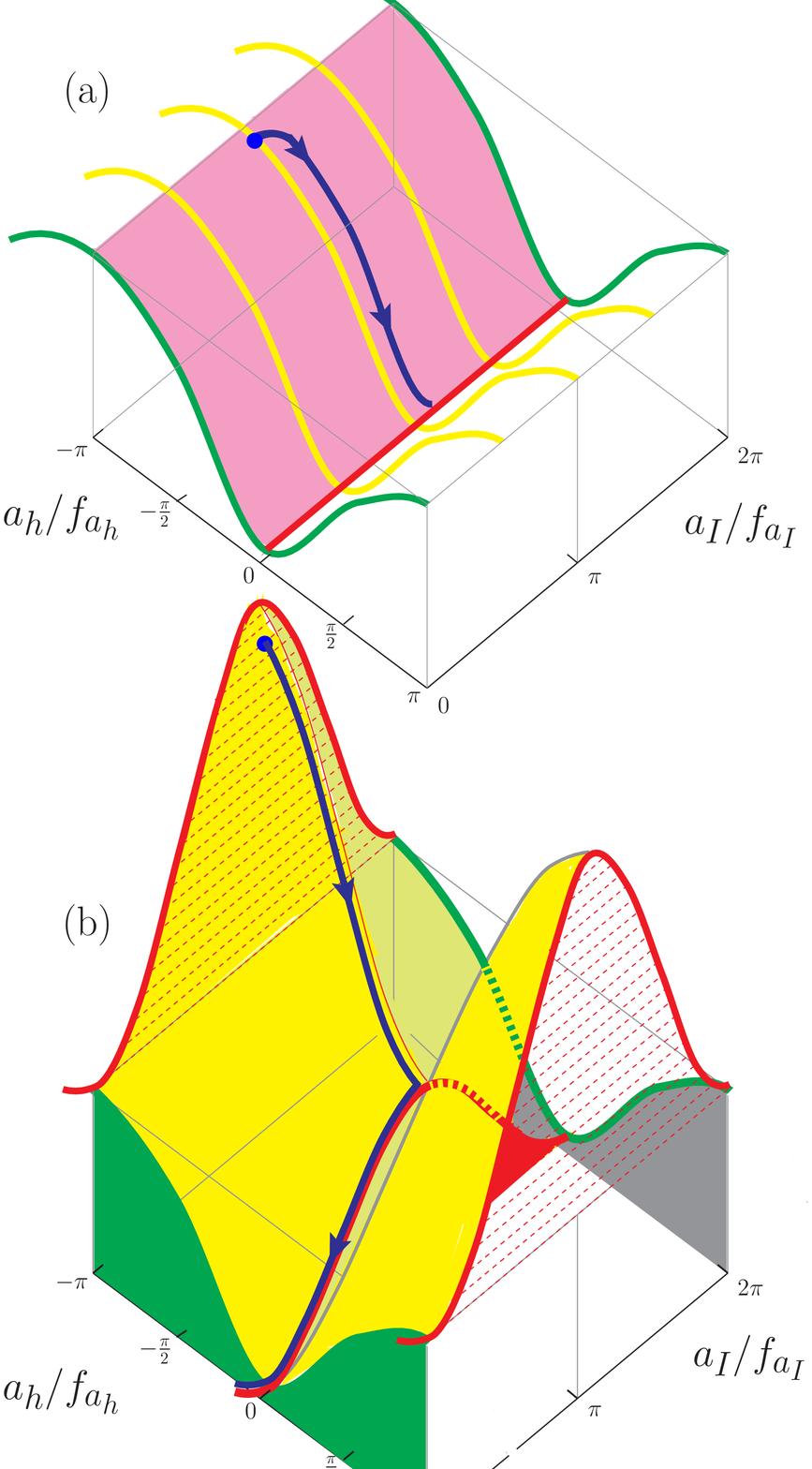}% This is an *.eps file
\end{center}
 \textbf{\refstepcounter{figure}\label{figKNP} Figure \arabic{figure}.}{  Two-flation. (a) The flat valley with one confining force is shown as the red line. The initial inflaton point rolls down by the heavy axion potential quickly to the flat valley, which is shown as the arrowed blue curve. (b) In the KNP model, two confining forces generate two mass eigenvalues, those of the heavy axion $a_{h}$ and the inflaton $a_{I}$.  The flat valley of (a) rises to the red valley and $f_{a_I}$ can be much larger than $f_{a_h}$. An inflation direction is shown as the blue arrowed-curve. }
\end{figure}

If there is only one confining force at the GUT scale, we can set $\Lambda_2=0$ in Eq. (\ref{eq:KNP2}). In this case, there exists a flat Goldstone boson direction as shown with the red valley in Fig. \ref{figKNP} \,(a). The heavy axion is
\begin{eqnarray}
a_h&=& \frac{(\alpha/\beta)a_1+ ( f_1/f_2) a_2}{\sqrt{(\alpha/\beta)^2 +( f_1/f_2)^2}}
\nonumber\\
&\equiv&\cos\theta\, a_1 +\sin\theta\, a_2 \propto \alpha\frac{a_1}{f_1}+\beta\frac{a_2}{f_2},
\label{eq:ah}
\end{eqnarray}
and the Goldstone boson direction is
\begin{eqnarray}
a_I&=&-\sin\theta\, a_1 +\cos\theta\, a_2,\label{eq:aI}
\end{eqnarray}
where
\begin{eqnarray}
 \cos\theta = \frac{\alpha/\beta}{\sqrt{(\alpha/\beta)^2 +( f_1/f_2)^2}},\\
 \sin\theta = \frac{f_1/f_2}{\sqrt{(\alpha/\beta)^2 +( f_1/f_2)^2}}.
\end{eqnarray}
Inverting the relations, we obtain
\begin{eqnarray}
 a_1&=&\cos\theta\, a_h - \sin\theta\,  a_I,  \label{eq:aone}\\
 a_2&=&\sin\theta\, a_h +\cos\theta\, a_I.\label{eq:atwo}
\end{eqnarray}
Therefore, the argument in the heavy axion potential is
\begin{eqnarray}
\frac{\alpha }{f_1}(\cos\theta\, a_h - \sin\theta\, a_I)+\frac{\beta }{f_2}(\sin\theta\, a_h +\cos\theta\, a_I )&&\nonumber\\
= \frac{\alpha }{f_1} \cos\theta\, a_h +\frac{\beta }{f_2} \sin\theta\, a_h
=\frac{  a_h}{f_1\cos\theta/\alpha}\, ,\quad\nonumber&&
\end{eqnarray}
with the condition, $\alpha\sin\theta/f_1= \beta\cos\theta/f_2$. Thus, the heavy axion decay constant is of order the GUT scale,
\begin{equation}
f_h=\frac{f_1\cos\theta}{\alpha}.
\end{equation}
The blue bullet field point of  Fig. \ref{figKNP}\,(a) quickly rolls along the blue path down to the red line vacuum. But the red line is a true Goldstone boson direction and there will be no inflation.

%%%%%%%%%%%%%%%%%%%%%%%%%%%%%%%%%%%%%%%%%%%%%%%%%%%%%%%%%%%%%%%%%%%%%%%%%%%%%%%%%%%%%%%

%%%%%%%%%%%%%%%%%%%%%%%%%%%%%%%%%%%%%%%%
\subsection{Two confining forces}\label{subsec:two}
If there are two confining forces, the situation is shown in Fig. \ref{figKNP} \,(b) \cite{KNP05}, with no Goldstone boson direction. In this case, we consider a $2\times 2$ mass matrix
\begin{equation}
\left(\begin{array}{cc}
   \frac{1}{f_1^2}\left({\alpha^2\Lambda^4_1}+ {\gamma^2\Lambda^4_2}
   \right), &\frac{1}{f_1f_2} (\alpha\beta\Lambda^4_1+ \gamma\delta\Lambda^4_2)
   \\[1em]
   \frac{1}{f_1f_2}(\alpha\beta\Lambda^4_1+ \gamma\delta\Lambda^4_2) , &
 \frac{1}{f_2^2}(\beta^2\Lambda^4_1+\delta^2\Lambda^4_2) \end{array}
\right).\nonumber\label{eq:Msquare}
 \end{equation}
Two eigenvalues of $M^2$ are
 \begin{eqnarray}
 m_{a_h}^2&=&\frac12(A+B),\\
 m_{a_I}^2&=&\frac12(A-B),\label{eq:mInf}
 \end{eqnarray}
 where
  \begin{eqnarray}
 A&=  \frac{\alpha^2\Lambda_1^4
+\gamma^2\Lambda_2^4}{f_1^2} +\frac{\beta^2\Lambda_1^4
+\delta^2\Lambda_2^4}{f_2^2}  \label{eq:A} \,, \label{eq:A}\\
B&=  \sqrt{A^2  -4(\alpha\delta -\beta\gamma)^2 \frac{\Lambda_1^4 \Lambda_2^4}{f_1^2 f_2^2} }\,. \label{eq:B}
 \end{eqnarray}
From Eqs. (\ref{eq:mInf},\ref{eq:A},\ref{eq:B}), note that the inflaton mass $m_{a_I}$ vanishes for $\alpha\delta=\beta\gamma$, which corresponds to an infinite $f_{a_I}$. Therefore, a large $f_{a_I}$ is possible for $\alpha\delta\approx \beta\gamma$. Let this approximation is described by a small number $\Delta$, \ie $\alpha\delta=\beta\gamma+\Delta$.
Then, the heavy axion and  inflaton masses are
\begin{eqnarray}
m_{a_h}^2&\simeq&\frac{\alpha^2\Lambda_1^4
+\gamma^2\Lambda_2^4}{f_1^2} +\frac{\beta^2\Lambda_1^4
+\delta^2\Lambda_2^4}{f_2^2},\nonumber\\[1em]
m_{a_I}^2 &\simeq& \frac{\Delta^2\Lambda_1^4\Lambda_2^4}{D}
  \end{eqnarray}
  where ${D=f_2^2(\alpha^2\Lambda_ 1^4 +\gamma^2\Lambda_2^4) +f_1^2(\beta^2\Lambda_1^4
+\delta^2\Lambda_2^4)}.$
For simplicity, let us discuss for $\Lambda_1=\Lambda_2=\Lambda$ and $f_1=f_2\equiv f$. Then, the masses are
\begin{eqnarray}
m_{a_h}^2&\simeq&(\alpha^2
+\beta^2 +\gamma^2+\delta^2 )\frac{\Lambda^4}{f^2}\,,\label{eq:mhapp} \\
 m_{a_I}^2&\simeq& \frac{\Lambda^4}{(\alpha^2 +\beta^2 +\gamma^2+\delta^2 )f^2/\Delta^2}\,,
\label{eq:mIapp}
 \end{eqnarray}
 from which we obtain
 \begin{equation}
 f_{a_I}=\frac{\sqrt{\alpha^2 +\beta^2 +\gamma^2+\delta^2 }f}{|\Delta|}.
 \label{eq:faI}
 \end{equation}
With the same order of $\alpha,\beta,\gamma,$ and $\delta$, the small number $\Delta$ can be O($1$) to realize $f_{a_I}\approx 100f$ if $\alpha,\beta,\gamma,\delta=$ O(50). Thus, the probability for $\Delta\approx 1$ to be realized is 1 out of $50\times 50$, \ie the large $f_{a_I}\approx 100f_{a_h}$ is possible in 0.04\% of random PQ quantum numbers $\alpha,\beta,\gamma$, and $\delta$ of O(50). However, the PQ quantum numbers $\alpha,\beta,\gamma$, and $\delta$ are not random priors, but are given definitely in a specific model. The case for a large $f_{a_I}$ is shown in Fig. \ref{figKNP}\,(b).

The flat valley of Fig. \ref{figKNP}\,(a) rises to the red valley of Fig. \ref{figKNP}\,(b) and $f_{a_I}$ can be $\approx 100f_{a_h}$ in a small region of the PQ quantum number space.
The blue bullet field point of  Fig. \ref{figKNP}\,(b) quickly rolls down along the blue path in the  $a_h$ direction down to the red valley. With a large decay constant of $a_I$, the red curve is very flat and provides the inflation path \cite{KNP05}, which is the 2-flation model.
The KNP inflation path is the blue arrowed-path on top of the red valley.

%%%%%%%%%%%%%%%%%%%%%%%%%%%%%%%%%%%%%%%%
\section{Number and sizes of non-Abelian gague groups}\label{sec:numberG}

The KNP 2-flation model has been generalized to N-flation models \cite{Nflation}.
The N-flation has adopted two merits of 2-flation, one that the decay constant is $\approx \sqrt2$ times larger and the other that  the maximum height of the potential is $\approx 2$ times larger. Namely, in the N-flation we expect that the decay constant can be $\approx \sqrt{N}$ times larger and the maximum height of the potential is $\approx N$ times larger. Then, from the highest point of the potential the denominators in the $\epsilon$ and $\eta$ calculation become $N$ times larger, making  $\epsilon$ and $\eta$ $N$ times smaller, and the decay constant is about $\sqrt{N}$ times larger. These merits are gradually diminished as the heavy axion paths shift directions as they roll down the hill.\footnote{See, for example, Ref. \cite{McDonald14}.}

In addition, in the N-flation the PQ quantum numbers are not tuned to large values of O(50).
However, an N-flation with a large $N$ suffers from the theoretical requirement of introducing $N (\gg 2)$ GUT scale non-Abelian gauge groups. In obtaining $N$, we must satisfy the SM phenomenology also. After realizing $\sin^2\theta^0_W=\frac38$ is needed \cite{Kim03}, non-prime orbifold compactification became popular since 2004 \cite{ChoiBook,Raby04}, and  $\sin^2\theta_W=\frac38$  is possible in many non-prime orbifold GUTs. In general, $\sin^2\theta_W=\frac38$ was not easy to be realized  in earlier orbifold models \cite{Ibanez87,Casas88,Ibanez93}. Successful SM construction have been obtained in ${\bf Z}_{12-I}$ \cite{KyaeZ12,KimJH,HuhKK09} and ${\bf Z}_{6-II}$ \cite{Lebedev07,Raby09,Ratz09}.
However, heterotic string models have not provided a useful moduli stabilization program, even though there exist some suggestions on stabilization of some moduli \cite{RaceTrack}.
Dynamical supersymmetry breaking \cite{susybreak} would also be another issue in a 2-flation model with a rank 16 gauge group.
In the heterotic string theory with level 1 construction, the sum of the ranks of gauge groups is 16 (or 22 in the Narain compactification \cite{Narain}.\footnote{For level greater than 1, it is possible to go beyond rank 16. We are aware of one example of construction at level 3 \cite{Tye96}, containing the SM gauge group (with suitable Higgs fields for breaking the gauge group) with three families of quarks and leptons. In principle, other string theories dual to the heterotic one with higher levels would also allow gauge groups whose rank is larger than 16. For instance, see
 \cite{Finst}.}
Out of rank 16, the SM uses 4 and rank 12 is left for the GUT scale confining gauge groups. If we use SU(4)'s for the N-flation, the maximal  $N$ is 3. This is hardly a case for the N-flation aims at. For $N=3$, the sum of the ranks of the GUT gauge groups is barely acceptable.
The probability for $\Delta\approx 1$, out of an approximate $10\times 10\times 10$ PQ quantum numbers (as we obtained Eq. (\ref{eq:faI} ) for the $N=2$ case) to obtain  $f_{a_I}\approx 100f$, is about 0.1\%. But this is from the PQ quantum numbers which are not really random priors.

In the D-brane construction of string theory, Ramond-Ramond (RR) charges of D-branes should be cancelled with proper orientifold p-planes ($O_p$-planes), which can be regarded as the fixed planes under a ${\bf Z}_2$ symmetry: anti-D branes can also compensate the RR charge of D-branes, but they hardly break SUSY, making the system unstable. The RR charge of an $O_p$-plane ($p=0,1,2,.., 9$), $Q_{Op}$ is given by $Q_{Op}=-2\cdot 2^{p-5}\times Q_{Dp}$, where $Q_{Dp}$ denotes the RR charge of a $Dp$ brane.\footnote{See, for example, Eq. (15) of Ref.  \cite{Giveon98}.}
[The maximum $Q_{Op}$ is $-32\, Q_{Dp}$.]
It can be cancelled by a stack of $N_c$ $Dp$ branes (the maximum number is $32\,Q_{Dp}$) parallel to an $Op$-plane, which yield a rank $N_c/2$ (the maximum number is 16) gauge group. Therefore, even in the D-brane construction, it is quite hard to obtain a gauge group whose rank is larger than 16.
%{\color{red}
%VEV of fermions possible?
%}

%%%%%%%%%%%%%%%%%%%%%%%%%%%%%%%%%%%%%%%%
\section{Conclusion}\label{sec:Conclusion}

The idea of natural inflation, using a GUT scale axion, has been extended to include $2, 3, \cdots, N$ axions. For the 2-flation, the PQ quantum numbers are almost degenerate, e.g. differing 1 out of 50. This almost degeneracy of the PQ quantum numbers can be relaxed by increasing $N$. In addition, the slow-roll parameters $\epsilon$ and $\eta$ can be reduced by a factor $1/N$. Models along this line can be constructed at field theory level.

However, in string compactification the number of non-Abelian gauge groups are restricted, which makes the realization of N-flation very difficult.
Most SM construction from string compactification used the level 1 construction in which case the rank is 16. Even if higher levels are assumed, the rank is 22. In any case, the rank cannot be of order 100.
Because of this difficulty of obtaining a large number of non-Abelian GUT scale gauge groups, the easiest realization of the trans-Planckian decay constant is the 2-flation. Nevertheless, it will be interesting to find out $N\ge 3$ non-Abelian GUT scale gauge groups from string compactification with the features satisfying the low energy SM phenomenology.

%%%%%%%%%%%%%%%%%%%%%%%%%%%%%%%%%%%%%%%%%%%%%%%%%%%%%%%%%%
\section*{Acknowledgement}
KYC is supported by the National Research Foundation (NRF) grant funded by the Korean Government (MEST) (No. 2011-0011083), JEK is supported in part by the NRF grant (No. 2005-0093841) and  by the IBS(IBS CA1310), and BK  is supported in part by the NRF grant (No. 2013R1A1A2006904).

%%%%%%%%%%%%%%%%%%%%%%%%%%%%%%%%%%%%%%%%
%%%%%%%%%%%%%%%%%%%%%%%%%%%%%%%%%%%%%%

\end{document}